\documentclass[aps,prd,11pt,nofootinbib]{revtex4-2}

\usepackage{amsmath,amssymb,amsfonts,mathtools,bm,mathrsfs}
\usepackage{booktabs}
\usepackage{microtype}
\usepackage{graphicx}
\usepackage[colorlinks=true,allcolors=blue]{hyperref}

\newcommand{\R}{\mathbb{R}}
\newcommand{\Z}{\mathbb{Z}}
\newcommand{\dd}{\mathrm{d}}

\newcommand{\diag}{\mathrm{diag}}
\newcommand{\EC}{\mathcal{E}_{\mathrm C}}
\newcommand{\ECzero}{\mathcal{E}_{\mathrm C}^{(0)}}
\newcommand{\ECsub}[1]{\mathcal{E}_{\mathrm C}^{#1}}
\newcommand{\V}{\mathcal{V}}
\newcommand{\kk}{\bm{k}}
\newcommand{\qq}{\bm{q}}
\newcommand{\nn}{\bm{n}}
\newcommand{\ellv}{\bm{\ell}}

\newcommand{\us}{\bm{u}_{s}}
\newcommand{\uc}{\bm{u}_{c}}
\newcommand{\un}{\bm{u}_{n}}
\newcommand{\Mmat}{M}
\newcommand{\Schur}{S}

\begin{document}

\title{Orientation-Dependent Casimir Energy and Generalized Torque in Lorentz-Violating Rectangular Compactifications}
\author{Zohreh Ashkipoor Gofreh}
\author{Reza Moazzemi}
\email{r.moazzemi@qom.ac.ir}
\affiliation{Department of Physics, Faculty of Science, University of Qom, Ghadir Blvd., Qom 371614-6611, I.R. Iran}

\begin{abstract}
We derive the renormalized Casimir energy of a free massive real scalar field on the rectangular compactification \(\mathbb{R}^{1,r}\times T^p\) in the presence of a constant CPT-even aether-like Lorentz-violating background. Allowing the spatial part of the background vector to have an arbitrary orientation relative to the compactification lattice, we characterize the resulting orientation-dependent geometric response of the vacuum. Integrating out the noncompact momenta reduces the effective spatial kinetic matrix to its Schur complement, and Poisson resummation yields a closed, exponentially convergent Bessel-function lattice sum within the stability domain. The resulting expression separates an orientation-independent kinetic normalization, \([1+\lambda u_\mu u^\mu]^{-1/2}\), from an orientation-dependent deformation of the image-lattice metric. Coordinate-aligned configurations are recovered as special cases, and the exact massless limit and the compact-direction pressures are obtained. For a two-dimensional compact subspace, we explicitly analyze the angular dependence of the vacuum energy and derive the associated generalized Casimir torque. These results establish how the relative orientation between a Lorentz-violating background and an anisotropic compactification lattice enters the renormalized vacuum energy and its directional response.

\end{abstract}

\maketitle

\section{Introduction}\label{sec:introduction}

The Casimir effect provides a particularly clean setting in which modifications of quantum vacuum fluctuations by global geometry and boundary conditions can be studied analytically. In quantum field theory, imposing finite-size conditions changes the spectrum of vacuum modes and produces a finite geometry-dependent contribution after the free-space part has been removed. For scalar fields in rectangular or hypercuboidal geometries, spectral sums and their analytic continuations provide a unified description of the dependence on field mass, dimensionality, and compactification lengths \cite{AmbjornWolfram1983,Bordag2009}.

Lorentz symmetry is a central spacetime symmetry of relativistic field theory. Nevertheless, effective-field-theory extensions allow small Lorentz-violating (LV) background structures while retaining observer covariance. A standard realization in the scalar sector is a CPT-even, aether-like derivative coupling to a constant background vector, of the form used below. Such terms can be viewed as a controlled deformation of the kinetic operator and have been studied extensively within the broader framework of the Standard-Model Extension (SME) \cite{ColladayKostelecky1997,ColladayKostelecky1998,Kostelecky2004}.

Vacuum energies are sensitive to LV modifications because they depend directly on the field spectrum. Earlier studies considered parallel-plate configurations with different orientations of a fixed background vector and different boundary conditions \cite{CruzBezerraPetrov2017,CruzBezerraMota2020,BezerraCruz2023}. More recently, a rectangular Dirichlet waveguide with a CPT-even aether-like vector was analyzed for four representative, coordinate-aligned configurations \cite{CruzBezerraMartin2026}. The latter work is particularly relevant to the present study because it shows explicitly how cavity geometry and background orientation enter the mode spectrum.

Related studies have examined Lorentz-violating Casimir physics in several other compactification settings. Reference \cite{Farias2026}, for example, considers a five-dimensional model with one additional compact dimension, quasiperiodic field conditions, and Neumann plates, whereas Ref. \cite{Cabral2026} investigates thermal and finite-size effects for a different Lorentz-violating tensor structure. These studies address spectral problems that are distinct from the one considered here. Moreover, an effective-metric and Schur-complement formulation has recently been discussed for Casimir energies in nonlinear scalar and electromagnetic theories \cite{Escobar2026}. Accordingly, the present work does not identify compactification or the Schur-complement reduction alone as its principal novelty.

Here we investigate the zero-temperature Casimir energy of a free massive real scalar field on \(\mathbb{R}^{1,r}\times T^p\), where the \(p\) spatial directions are periodically identified with independently specified compactification lengths. Unlike the rectangular Dirichlet-waveguide problem of Ref. \cite{CruzBezerraMartin2026}, our setup has a periodic spectrum and allows the spatial projection of the constant CPT-even aether-like background vector to have an arbitrary orientation relative to the compactification lattice. The technical core of our analysis is that the Lorentz-violating deformation modifies the compactification geometry through an effective positive-definite spatial quadratic form. Integrating out the noncompact momenta reduces this form to the Schur complement of the noncompact kinetic block, while Poisson resummation yields a closed, exponentially convergent Bessel-function lattice representation of the renormalized Casimir energy. The resulting formulation separates an orientation-independent kinetic normalization from an orientation-dependent deformation of the image-lattice distances. We further derive the massless limit and the compact-direction pressures. For a two-dimensional compact subspace, we explicitly determine the angular dependence of the vacuum energy and the associated generalized Casimir torque. The principal contribution of this work is thus a unified, all-dimensional treatment of arbitrary background orientation and independently sized periodic compact directions, together with an explicit characterization of their geometric anisotropy.

The paper is organized as follows: Subsection~\ref{sec:aw-viewpoint} recalls the Ambj\o rn--Wolfram spectral framework and explains how the Lorentz-violating deformation enters it. In Section~\ref{sec:lv-model} we briefly introduce the scalar-field model and derive its modified dispersion relation and stability conditions. Section~\ref{sec:compactification} formulates the rectangular periodic compactification and the Schur-complement reduction and presents the Poisson-resummed compact spectrum. The master formula for the renormalized Casimir energy is derived in Section~\ref{sec:casimir-master}. Section~\ref{sec:axis-aligned} recovers the principal axis-aligned limits. In Section~\ref{sec:generic-orientation} we develop the generic-orientation result, including the weak-LV expansion and an explicit angular example. Section~\ref{sec:pressures-torque} derives the compact-direction pressures and the orientation-induced generalized torque. Section~\ref{sec:massless-checks} presents the massless limit and normalization checks, and also compares the periodic compactification with the Dirichlet rectangular-waveguide problem. In Section~\ref{sec:discussion} we discuss the physical interpretation and limitations of the results, and Section~\ref{sec:conclusion} summarizes the main conclusions. Appendix~\ref{app:technical-derivation} provides details of the Gaussian integrations and Poisson resummation.

\subsection{The Ambj\o rn--Wolfram viewpoint and its deformation}\label{sec:aw-viewpoint}
A major strength of the Ambj\o rn--Wolfram construction is that the finite-size contribution is organized at the level of the spectrum rather than tied to one special geometry. For a rectangular compactification, the Lorentz-invariant problem can be encoded by the compactification matrix $R=\diag(a_1,\ldots,a_p)$ and the image-lattice distance
\begin{equation}
    \left(L_{\ell}^{(0)}\right)^2=\ellv^{\mathsf T}R^2\ellv.
    \label{eq:AWdistance}
\end{equation}
After removing the zero-image sector, the corresponding Bessel representation is
\begin{equation}
    \ECzero
    =-\frac{\V}{2\sqrt{\pi}(4\pi)^{d/2}}
    \sum_{\ellv\in\Z^p\setminus\{0\}}
    \left(\frac{2m}{L_{\ell}^{(0)}}\right)^{\nu}
    K_{\nu}\!\left(mL_{\ell}^{(0)}\right),
    \qquad \nu=\frac{d+1}{2}.
    \label{eq:AWresult}
\end{equation}
The usefulness of this organization is not merely the existence of a closed form. The geometry of the compactification is separated from the analytic continuation of the spectral sum, while the zero-image sector is identified as the bulk contribution. This makes the construction particularly well suited to controlled deformations of the kinetic operator.

The present work preserves this organization. Lorentz violation replaces the ordinary spatial kinetic form by a positive-definite matrix $\Mmat$. Once the noncompact variables are integrated out, the replacement of the Ambj\o rn--Wolfram image-lattice metric is
\begin{equation}
    R^2 \longrightarrow R\Schur^{-1}R,
    \qquad
    \text{together with the normalization}
    \quad [1+\lambda u_\mu u^\mu]^{-1/2}.
    \label{eq:AWdeformation}
\end{equation}
Thus the new ingredient is a deformation of the spectral geometry underlying the known construction, rather than a replacement of the Casimir machinery itself.

The main result is a compact expression for the renormalized Casimir energy in which the orientation dependence is encoded in the metric of the Poisson-resummed image lattice generated by a Schur complement. In particular, the LV background produces two conceptually different effects: an orientation-independent normalization of the vacuum energy and an orientation-sensitive deformation of the effective compactification lengths. This makes it possible to distinguish geometries in which the LV correction is only multiplicative from those in which the relative angle between the background and the compactification lattice becomes physically relevant.

\section{Lorentz-violating scalar model}\label{sec:lv-model}
We work in flat $(d+1)$-dimensional Minkowski spacetime with metric signature
\begin{equation}
    \eta_{\mu\nu}=\diag(+,-,\ldots,-).
\end{equation}
The real scalar field is described by the Lagrangian density
\begin{equation}
    \mathcal{L}
    =\frac12\left[
    \partial_\mu\phi\,\partial^\mu\phi
    -m^2\phi^2
    +\lambda\left(u^\mu\partial_\mu\phi\right)^2
    \right],
    \label{eq:lagrangian}
\end{equation}
where $m$ is the scalar mass, $\lambda$ is a dimensionless LV coefficient, and $u^\mu$ is a constant nondynamical $(d+1)$-vector. The model is the CPT-even, aether-like scalar structure commonly used in studies of Lorentz violation \cite{CruzBezerraPetrov2017,Kostelecky2004}. The constant-vector assumption preserves translational invariance.

The equation of motion is
\begin{equation}
    \left[\Box+m^2+\lambda(u\!\cdot\!\partial)^2\right]\phi(x)=0,
    \qquad
    u\!\cdot\!\partial\equiv u^\mu\partial_\mu.
    \label{eq:eom}
\end{equation}
It is useful to separate the temporal and spatial parts of the background,
\begin{equation}
    u^\mu=(u^0,\us),
    \qquad
    \kk\in\R^d,
\end{equation}
and to use the plane-wave convention
\begin{equation}
    \phi(x)=\phi_0\exp[-i\omega t+i\kk\cdot\bm{x}].
\end{equation}
Defining
\begin{equation}
    q=\us\cdot\kk,
    \qquad
    \alpha=1+\lambda(u^0)^2,
    \qquad
    \rho=\frac{\lambda}{\alpha},
    \label{eq:alpharho}
\end{equation}
the modified Klein--Gordon equation becomes
\begin{equation}
    \alpha\omega^2-2\lambda u^0q\,\omega
    -\left(\kk^2+m^2-\lambda q^2\right)=0.
    \label{eq:dispquad}
\end{equation}
The positive-frequency branch can therefore be written as
\begin{equation}
    \omega(\kk)
    =\frac{\lambda u^0q}{\alpha}
    +\frac{1}{\sqrt{\alpha}}
    \sqrt{\kk^{\mathsf T}\Mmat\kk+m^2},
    \label{eq:dispersion}
\end{equation}
where the effective spatial kinetic matrix is
\begin{equation}
    \Mmat=I_d-\rho\,\us\us^{\mathsf T}.
    \label{eq:Mmatrix}
\end{equation}
The first term in Eq.~\eqref{eq:dispersion} is odd under $\kk\to-\kk$. For the symmetric mode lattice and symmetric integration domains used below, it cancels identically from the vacuum-energy sum. The nontrivial LV dependence is consequently encoded in the quadratic form $\kk^{\mathsf T}\Mmat\kk$.

For a real scalar field with a stable oscillatory spectrum, it is sufficient to require
\begin{equation}
    \alpha>0,
    \qquad
    \det\Mmat=1-\rho|\us|^2>0.
    \label{eq:stability}
\end{equation}
The second condition can also be written as
\begin{equation}
    1+\lambda u_\mu u^\mu>0,
    \qquad
    u_\mu u^\mu=(u^0)^2-|\us|^2.
    \label{eq:stability_invariant}
\end{equation}
Within this domain, $\Mmat$ is positive definite. A useful exact identity that will simplify the final answer is
\begin{equation}
    \alpha\det\Mmat=1+\lambda u_\mu u^\mu.
    \label{eq:alphadet}
\end{equation}

\section{Rectangular compactification}\label{sec:compactification}
We compactify $p$ spatial directions with independent periodic identifications
\begin{equation}
    x_i\sim x_i+a_i,
    \qquad i=1,\ldots,p,
\end{equation}
and leave the remaining $r=d-p$ spatial directions noncompact. Introduce the diagonal matrix
\begin{equation}
    R=\diag(a_1,\ldots,a_p),
    \qquad
    \V=\det R=\prod_{i=1}^{p}a_i.
    \label{eq:Rmatrix}
\end{equation}
The compact momenta are quantized as
\begin{equation}
    \kk_c=2\pi R^{-1}\nn,
    \qquad
    \nn\in\Z^p,
    \label{eq:compactmom}
\end{equation}
while the noncompact momenta $\kk_n$ remain continuous.

Split the spatial momentum and the effective kinetic matrix according to compact and noncompact subspaces,
\begin{equation}
    \kk=\begin{pmatrix}\kk_c\\ \kk_n\end{pmatrix},
    \qquad
    \Mmat=
    \begin{pmatrix}
      A & B\\
      B^{\mathsf T} & C
    \end{pmatrix}.
    \label{eq:blockM}
\end{equation}
The quadratic form can be completed in the noncompact variables:
\begin{align}
    \kk^{\mathsf T}\Mmat\kk
    &=(\kk_n+C^{-1}B^{\mathsf T}\kk_c)^{\mathsf T}
    C(\kk_n+C^{-1}B^{\mathsf T}\kk_c)
    +\kk_c^{\mathsf T}\Schur\kk_c,
    \\[-1mm]
    \Schur&=A-BC^{-1}B^{\mathsf T}.
    \label{eq:schur}
\end{align}
Thus $\Schur$ is the Schur complement of the noncompact kinetic block. Since $\Mmat$ is positive definite within the stability domain, both $C$ and $\Schur$ are positive definite. For $r=0$ there is no noncompact block; in that case the Schur-complement step is absent and one simply sets $\Schur=\Mmat$.

Writing the spatial background as $\us=(\uc,\un)$ gives the explicit block matrices
\begin{equation}
    C=I_r-\rho\,\un\un^{\mathsf T},
    \qquad
    \det C=1-\rho|\un|^2,
    \label{eq:Cblock}
\end{equation}
and
\begin{equation}
    \Schur
    =I_p-\frac{\rho}{1-\rho|\un|^2}\,\uc\uc^{\mathsf T}.
    \label{eq:Sexpl}
\end{equation}
The compact quadratic form is therefore
\begin{equation}
    \kk_c^{\mathsf T}\Schur\kk_c
    =\nn^{\mathsf T}Q\nn,
    \qquad
    Q=4\pi^2R^{-1}\Schur R^{-1}.
    \label{eq:Qform}
\end{equation}
For a generic orientation of $\uc$, the matrix $Q$ is non-diagonal even when the geometric compactification itself is rectangular. This is the key structural difference between an axis-aligned calculation and the general problem.\footnote{The Schur-complement step is not restricted to the rank-one kinetic deformation $\lambda u_\mu u_\nu$ used here. For a more general constant symmetric Lorentz-violating kinetic coefficient, producing a quadratic momentum term of the form $k^\mu c_{\mu\nu}k^\nu$, the same block-matrix reduction applies whenever the effective spatial kinetic matrix is positive definite. The corresponding dispersion relation, stability conditions, and determinant prefactor must still be derived for the chosen tensor structure. This provides a natural extension to the tensor-coefficient language of the SME \cite{ColladayKostelecky1998,Kostelecky2004}.}

\subsection{Poisson-resummed form of the compact spectrum}\label{subsec:poisson-spectrum}
The zero-point contribution per unit volume of the noncompact directions is formally
\begin{equation}
    \ECsub{\rm formal}
    =\frac{1}{2\sqrt{\alpha}}
    \sum_{\nn\in\Z^p}
    \int\frac{\dd^r k_n}{(2\pi)^r}
    \sqrt{\kk^{\mathsf T}\Mmat\kk+m^2}.
    \label{eq:formalE}
\end{equation}
The extensive uncompactified vacuum contribution is the zero-winding sector after Poisson resummation and will be subtracted.

For later use, the multidimensional Poisson identity appropriate to the present quadratic form is
\begin{equation}
\sum_{\nn\in\Z^p}
\exp\!\left[-4\pi^2 t\,\nn^{\mathsf T}R^{-1}\Schur R^{-1}\nn\right]
=\frac{\V}{(4\pi t)^{p/2}\sqrt{\det\Schur}}
\sum_{\ellv\in\Z^p}
\exp\!\left[-\frac{1}{4t}\ellv^{\mathsf T}R\Schur^{-1}R\ellv\right].
\label{eq:poisson}
\end{equation}
The quantity
\begin{equation}
    L_\ell^2
    =\ellv^{\mathsf T}R\Schur^{-1}R\ellv
    \label{eq:dualLength}
\end{equation}
plays the role of an LV-deformed Poisson-resummed image-lattice distance.

\section{Casimir energy: master formula}\label{sec:casimir-master}
We use the proper-time representation
\begin{equation}
    \sqrt{X}
    =-\frac{1}{2\sqrt{\pi}}
    \int_0^\infty\dd t\,t^{-3/2}e^{-tX},
    \qquad X>0,
    \label{eq:proper}
\end{equation}
and perform the Gaussian integral over the $r$ noncompact momenta. After Poisson resummation, Eq.~\eqref{eq:formalE} becomes
\begin{equation}
    \EC
    =-\frac{\V}{4\sqrt{\pi}(4\pi)^{d/2}
    \sqrt{\alpha\det\Mmat}}
    \sum_{\ellv\in\Z^p\setminus\{0\}}
    \int_0^\infty\dd t\,t^{-(d+3)/2}
    \exp\!\left[-m^2t-\frac{L_\ell^2}{4t}\right].
    \label{eq:properfinite}
\end{equation}
The omitted $\ellv=0$ term is the bulk vacuum energy and contains no information about the finite compactification scales.

The remaining integral can be evaluated with
\begin{equation}
    \int_0^\infty\dd t\,t^{-\nu-1}e^{-\beta t-\gamma/t}
    =2\left(\frac{\gamma}{\beta}\right)^{-\nu/2}
    K_\nu\!\left(2\sqrt{\beta\gamma}\right),
    \qquad \Re\beta,\Re\gamma>0,
    \label{eq:besselidentity}
\end{equation}
where $\nu=(d+1)/2$. We obtain the central result
\begin{equation}
 {
\EC
=-\frac{\V}{2\sqrt{\pi}(4\pi)^{d/2}}
\frac{1}{\sqrt{1+\lambda u_\mu u^\mu}}
\sum_{\ellv\in\Z^p\setminus\{0\}}
\left(\frac{2m}{L_\ell}\right)^{\frac{d+1}{2}}
K_{\frac{d+1}{2}}(mL_\ell)
}
\label{eq:master}
\end{equation}
where we used Eq.~\eqref{eq:alphadet}. For $r=0$ the quantity $\EC$ is the total Casimir energy, whereas for $r>0$ it is the energy per unit $r$-dimensional noncompact volume.

Equation~\eqref{eq:master} is the all-dimensional result for the periodic rectangular compactification. It exposes two distinct LV effects. First, the factor $[1+\lambda u_\mu u^\mu]^{-1/2}$ is an overall kinetic normalization. Second, the length $L_\ell$ depends on the orientation of the spatial projection of $u^\mu$ relative to the compactification lattice. The latter is the genuinely geometric effect that is absent in the axis-aligned normalization-only cases.

\begin{figure}[t]
    \centering
    \includegraphics[width=0.78\linewidth]{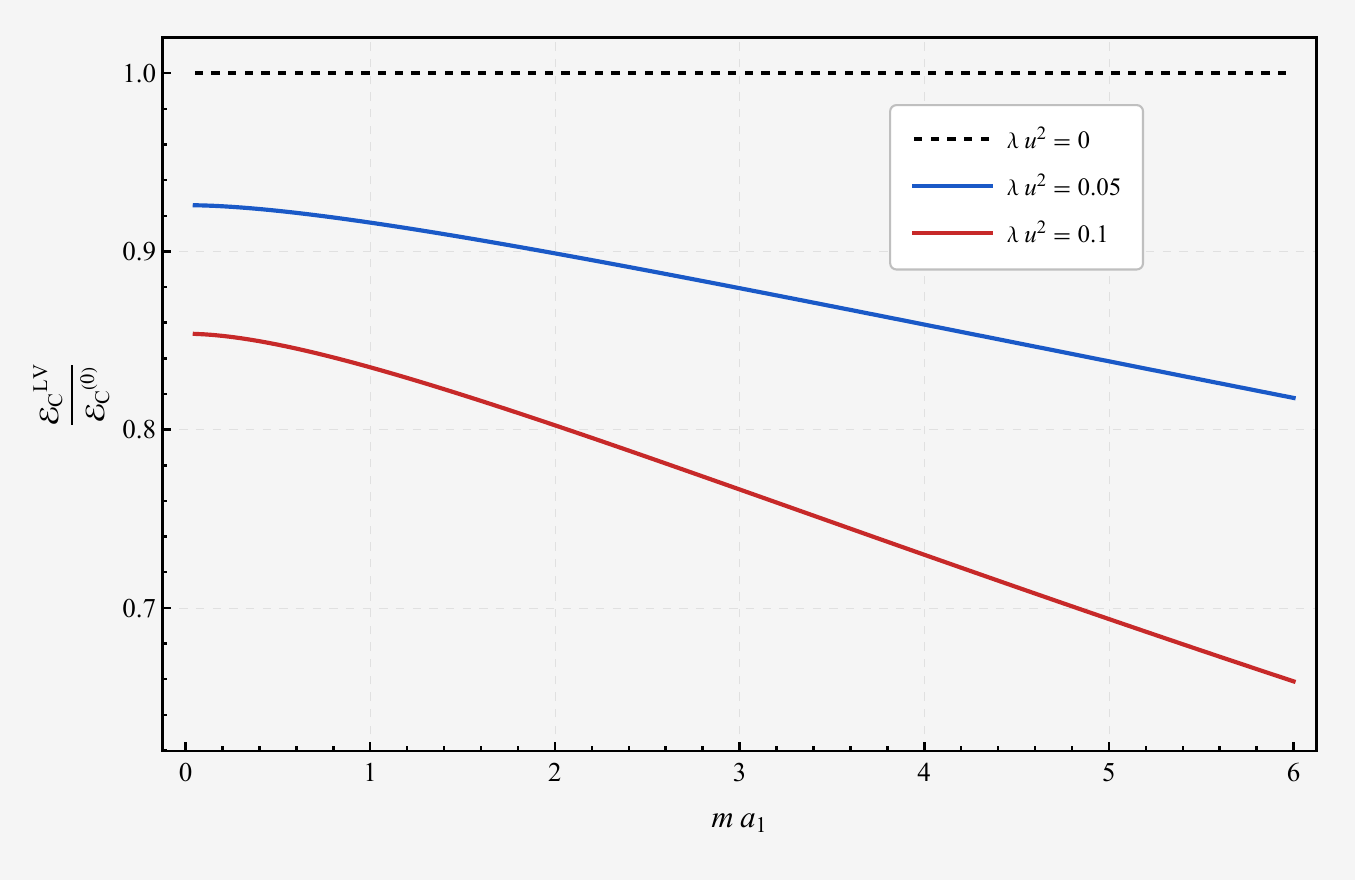}
    \caption{Massive Casimir energy per unit noncompact volume, normalized to the Lorentz-invariant value, as a function of $ma_1$ for $d=3$, $p=1$, and a purely spatial background aligned with the compact direction. The curves correspond to $\lambda u^2=0$, $0.05$, and $0.10$.}
    \label{fig:masslambda}
\end{figure}

\begin{figure}[t]
    \centering
    \includegraphics[width=0.78\linewidth]{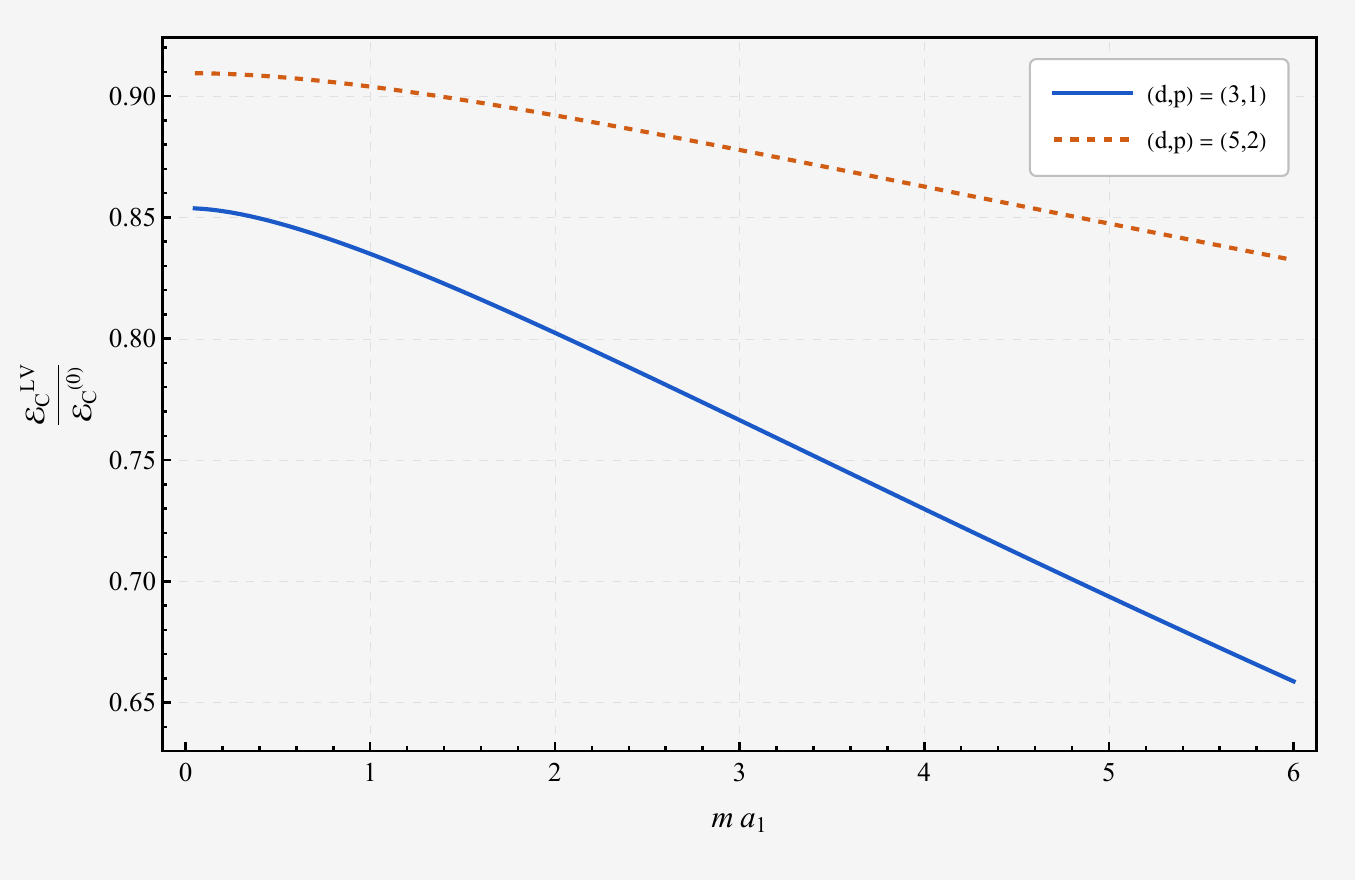}
    \caption{Normalized massive Casimir energy versus $ma_1$ for $(d,p)=(3,1)$ and $(5,2)$ at fixed $\lambda u^2=0.1$, with equal compact lengths and a background aligned with one compact axis.}
    \label{fig:dp}
\end{figure}

To illustrate these effects, Fig.~\ref{fig:masslambda} shows the massive Casimir energy normalized to its Lorentz-invariant value for a background aligned with a compact direction. Figure~\ref{fig:dp} compares two choices of total dimension and compactification rank at fixed LV strength. 

\begin{table}[t]
\centering
\caption{Principal limits of the master formula}
\label{tab:limits}
\begin{tabular}{p{0.27\linewidth} p{0.63\linewidth}}
\toprule
Configuration & Effect on the Casimir energy \\
\midrule
Purely timelike background & Pure multiplicative factor $[1+\lambda(u^0)^2]^{-1/2}$ \\
Purely spatial background in a compact direction & Equivalent to a rescaling of the corresponding compact length $a_i\to a_i/\sqrt{1-\lambda u^2}$ \\
Purely spatial background with no compact projection & Pure multiplicative factor $(1-\lambda|\un|^2)^{-1/2}$ \\
Generic spatial orientation & Anisotropic deformation of the Poisson-resummed image lattice through $L_\ell$ \\
\bottomrule
\end{tabular}
\end{table}

\section{Recovery of the axis-aligned cases}\label{sec:axis-aligned}
\subsection{Purely timelike background}\label{subsec:timelike-background}
For
\begin{equation}
    u^\mu=(u^0,\bm{0}),
\end{equation}
one has $\Mmat=I_d$, $\Schur=I_p$, and
\begin{equation}
    \left(L_\ell^{(0)}\right)^2
    =\sum_{i=1}^{p}(a_i\ell_i)^2.
    \label{eq:L0}
\end{equation}
Consequently,
\begin{equation}
    \ECsub{\rm timelike}
    =\frac{1}{\sqrt{1+\lambda(u^0)^2}}\,\ECzero.
    \label{eq:timelike}
\end{equation}
Thus a purely timelike background rescales the normalization of the vacuum energy but does not alter the compactification geometry.

\subsection{Spacelike background along a compact direction}\label{subsec:compact-axis}
Let the background be purely spatial and aligned with compact direction $x_i$:
\begin{equation}
    u^\mu=(0,0,\ldots,u,\ldots,0).
\end{equation}
Then
\begin{equation}
    \det\Mmat=1-\lambda u^2,
    \qquad
    \Schur=\diag(1,\ldots,1-\lambda u^2,\ldots,1),
\end{equation}
and hence
\begin{equation}
    L_\ell^2
    =\sum_{j\ne i}(a_j\ell_j)^2
    +\frac{(a_i\ell_i)^2}{1-\lambda u^2}.
    \label{eq:compactL}
\end{equation}
This can be interpreted exactly as the Lorentz-invariant rectangular result with an effective length
\begin{equation}
    a_i^{\rm eff}=\frac{a_i}{\sqrt{1-\lambda u^2}}.
    \label{eq:aeff}
\end{equation}
The LV correction in this case is therefore not merely a prefactor: it changes the geometric scale sampled by the vacuum modes.

\subsection{Purely spatial background with no compact projection}\label{subsec:noncompact-axis}
Suppose $\uc=0$ but $\un\ne0$. Then $\Schur=I_p$ and the only modification is a Jacobian from the noncompact momentum integral,
\begin{equation}
    \ECsub{\rm noncompact}
    =\frac{1}{\sqrt{1-\lambda|\un|^2}}\,\ECzero.
    \label{eq:noncompact}
\end{equation}
This factor is independent of $r=d-p$. The reason is structural: after a rotation inside the noncompact subspace, the background can be aligned with one momentum component, and the required change of variables is only
\begin{equation}
    q_1=\sqrt{1-\lambda|\un|^2}\,k_1,
    \qquad
    q_j=k_j\quad (j=2,\ldots,r).
    \label{eq:noncompact_rescaling}
\end{equation}
Thus only one integration measure contributes the square-root Jacobian. A power depending on the total number of noncompact dimensions would instead arise from rescaling all continuous momentum components, which is not the transformation associated with a background supported in a single noncompact eigendirection. Only when $\uc\ne0$ does the compactification lattice break the continuous rotational freedom and make the relative orientation geometrically observable.

\section{Generic orientation and the induced quadratic form}\label{sec:generic-orientation}
For a generic decomposition
\begin{equation}
    \us=(\uc,\un),
\end{equation}
the inverse of the Schur complement gives
\begin{equation}
    \Schur^{-1}
    =I_p+\frac{\rho}{1-\rho|\us|^2}\,\uc\uc^{\mathsf T}.
\end{equation}
Substitution into Eq.~\eqref{eq:dualLength} yields the explicit Poisson-resummed image-lattice distance
\begin{equation}
     {
    L_\ell^2
    =\ellv^{\mathsf T}R^2\ellv
    +\frac{\rho}{1-\rho|\us|^2}
    \left(\uc^{\mathsf T}R\ellv\right)^2
    }
    \label{eq:Lexplicit}
\end{equation}
with $\rho=\lambda/\alpha$.

Equation~\eqref{eq:Lexplicit} makes the orientation dependence particularly transparent. The LV correction is controlled by the projection of the background onto the corresponding image-lattice vector $R\ellv$. Consequently, the effect depends not only on the magnitude of $u^\mu$ but also on the relative orientation between the background and the rectangular lattice.

\subsection{Weak-LV expansion}\label{subsec:weak-lv}
For $|\lambda|\ll1$ one finds
\begin{equation}
    L_\ell^2
    =\left(L_\ell^{(0)}\right)^2
    +\lambda\left(\uc^{\mathsf T}R\ellv\right)^2
    +\mathcal O(\lambda^2),
    \label{eq:Lexpansion}
\end{equation}
and
\begin{equation}
    \frac{1}{\sqrt{1+\lambda u_\mu u^\mu}}
    =1-\frac{\lambda}{2}u_\mu u^\mu+
    \mathcal O(\lambda^2).
    \label{eq:prefexpansion}
\end{equation}
The leading correction therefore separates naturally into an invariant normalization part and an orientation-sensitive lattice deformation. If $\uc=0$, the geometric correction vanishes and the exact result reduces to the single-Jacobian factor in Eq.~\eqref{eq:noncompact}, even when several noncompact components of $u^\mu$ are nonzero.

\subsection{Explicit angular dependence for a two-dimensional compact subspace}\label{subsec:angular-example}
To display the geometric content without introducing unnecessary dimensional complexity, consider $d=3$, $p=2$ and a purely spatial background confined to the compact plane,
\begin{equation}
    \uc=u(\cos\theta,\sin\theta),
    \qquad
    \un=0.
    \label{eq:angleu}
\end{equation}
For the massless field, the corresponding expression follows by taking the limit $m\to0^+$ directly in Eq.~\eqref{eq:master}; the general massless formula is derived explicitly in Sec.~\ref{sec:massless-checks}. For a square compactification, $a_1=a_2$, the spectrum is invariant under the discrete square-lattice symmetry and the angular dependence is correspondingly periodic. For a rectangular compactification, $a_1\ne a_2$, this discrete symmetry is reduced and the angular modulation becomes stronger.

To expose the leading angular dependence analytically, define $\epsilon\equiv\lambda u^2$ and the Lorentz-invariant lattice length
\begin{equation}
    \left(L_\ell^{(0)}\right)^2=(a_1\ell_1)^2+(a_2\ell_2)^2.
\end{equation}
For $|\epsilon|\ll1$, expansion of the massless energy gives
\begin{equation}
    \EC(\theta)=-\frac{V_2}{2\pi^2}\left[S_0+\epsilon\left(\frac{S_0}{2}-2S_x\cos^2\theta-2S_y\sin^2\theta\right)\right]+\mathcal O(\epsilon^2),
    \label{eq:weak-angle-energy}
\end{equation}
where $V_2=a_1a_2$ and the sums contain only the undeformed lattice lengths,
\begin{align}
    S_0&=\sum_{\ellv\in\Z^2\setminus\{0\}}\frac{1}{(L_\ell^{(0)})^4},\nonumber\\
    S_x&=\sum_{\ellv\in\Z^2\setminus\{0\}}\frac{(a_1\ell_1)^2}{(L_\ell^{(0)})^6},\qquad
    S_y=\sum_{\ellv\in\Z^2\setminus\{0\}}\frac{(a_2\ell_2)^2}{(L_\ell^{(0)})^6}.
    \label{eq:weak-angle-sums}
\end{align}
The mixed term proportional to $\sin\theta\cos\theta$ vanishes by reflection symmetry of the lattice sum, and $S_x+S_y=S_0$. Consequently, the normalized angular dependence is
\begin{equation}
    \frac{\EC(\theta)}{\EC(0)}=1-2\epsilon\frac{S_y-S_x}{S_0}\sin^2\theta+\mathcal O(\epsilon^2).
    \label{eq:weak-angle-ratio}
\end{equation}
This makes the role of geometric anisotropy explicit: for a square lattice, $S_x=S_y=S_0/2$, so the angular modulation of the normalized energy vanishes at first order in $\epsilon$; for a rectangular lattice, $S_x$ and $S_y$ need not coincide.

\begin{figure}[t]
    \centering
    \includegraphics[width=0.78\linewidth]{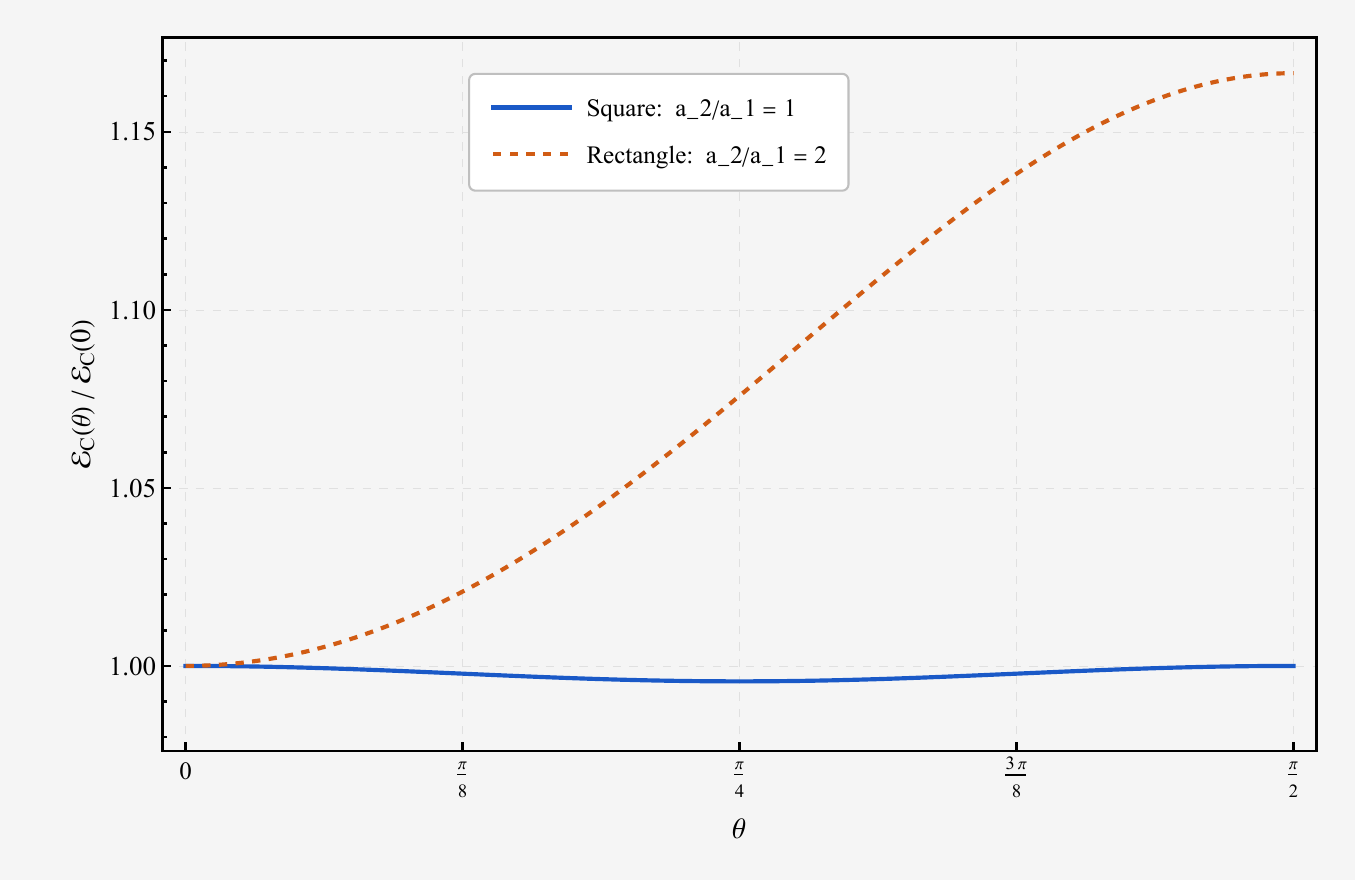}
    \caption{Angular dependence of the massless Casimir energy for $d=3$, $p=2$, and $\lambda u^2=0.1$, with the background vector lying in the compact plane. The energy is normalized to its value at $\theta=0$. The curves compare a square compactification, $a_2/a_1=1$, with a rectangular compactification, $a_2/a_1=2$.}
    \label{fig:angular}
\end{figure}

As an illustration, Fig.~\ref{fig:angular} shows the ratio $\EC(\theta)/\EC(0)$ for $\lambda u^2=0.1$ and the two choices $a_2/a_1=1$ and $a_2/a_1=2$. The plotted values were obtained by direct numerical summation of the massless lattice expression with the symmetric cutoff $|\ell_i|\le100$, $\ellv\ne0$; increasing the cutoff from $60$ to $100$ changes the displayed ratio near $\theta=\pi/4$ by less than $5\times10^{-6}$ for the square case and about $2\times10^{-6}$ for the rectangular case.

The example demonstrates the central physical consequence of the general formulation: once the compactification lengths are inequivalent, the vacuum energy becomes an orientation-sensitive quantity and therefore supports the torque defined in Eq.~\eqref{eq:torque}. This effect disappears in the limit $\lambda\to0$ and is distinct from the overall normalization induced by the Lorentz-violating kinetic term.

\section{Casimir pressures and the orientation-induced torque}\label{sec:pressures-torque}
Let $V_p=\prod_{j=1}^p a_j$ denote the compactification volume, and let $E=V_r\EC$ be the total vacuum energy when a finite noncompact volume $V_r$ is used. The pressure conjugate to a change of the compact length $a_i$ is
\begin{equation}
    P_i=-\frac{a_i}{V_p}\frac{\partial \EC}{\partial a_i},
    \qquad i=1,\ldots,p.
    \label{eq:pressure}
\end{equation}
It is useful to introduce
\begin{equation}
    \nu=\frac{d+1}{2},
    \qquad
    \Delta=1+\lambda u_\mu u^\mu,
    \qquad
    \mathscr{C}_d=\frac{1}{2\sqrt{\pi}(4\pi)^{d/2}\sqrt{\Delta}},
    \label{eq:Cdef}
\end{equation}
and
\begin{equation}
    f_\nu(L)=\left(\frac{2m}{L}\right)^\nu K_\nu(mL).
\end{equation}
Then Eq.~\eqref{eq:master} reads $\EC=-\mathscr{C}_d V_p\sum_{\ellv\ne0}f_\nu(L_\ell)$. Using
\begin{equation}
    \frac{\partial f_\nu}{\partial L}
    =-m(2m)^\nu L^{-\nu}K_{\nu+1}(mL),
    \label{eq:dfdL}
\end{equation}
the directional pressures can be written explicitly as
\begin{equation}
 {
    P_i=\mathscr{C}_d\sum_{\ellv\ne0}
    \left[
    f_\nu(L_\ell)
    -\frac{m}{2}(2m)^\nu L_\ell^{-\nu-1}
    K_{\nu+1}(mL_\ell)
    D_{i\ell}
    \right],
    }
\label{eq:explicitpressure}
\end{equation}
where
\begin{equation}
    D_{i\ell}
    \equiv a_i\frac{\partial L_\ell^2}{\partial a_i}
    =2(a_i\ell_i)^2
    +2\chi\,(u_{c,i}a_i\ell_i)\left(\uc^{\mathsf T}R\ellv\right),
    \qquad
    \chi\equiv\frac{\rho}{1-\rho|\us|^2}.
    \label{eq:Diell}
\end{equation}
This makes the mechanical content of the spectral result explicit: when the background has a compact projection, the same orientation-sensitive quadratic form that deforms the image lattice also modifies the directional vacuum pressures. For unequal compact lengths, there is in general no symmetry requirement that $P_i=P_j$.

\subsection{Casimir torque in a two-dimensional compact subspace}\label{subsec:torque-angular}
A particularly transparent observable is the torque associated with rotating the background relative to an anisotropic compactification. In the two-dimensional compact plane, take
\begin{equation}
	\uc=u(\cos\theta,\sin\theta),
	\qquad \un=0,
	\qquad d=3,
	\qquad p=2.
	\label{eq:torque_u}
\end{equation}
Define
\begin{align}
	X_{\ell} &=a_1\ell_1\cos\theta+a_2\ell_2\sin\theta,\\
	Y_{\ell} &=-a_1\ell_1\sin\theta+a_2\ell_2\cos\theta.
\end{align}
Then
\begin{equation}
	L_\ell^2=(a_1\ell_1)^2+(a_2\ell_2)^2+\chi u^2X_\ell^2,
	\qquad
	\chi=\frac{\lambda}{1-\lambda u^2}.
	\label{eq:Langle}
\end{equation}
The torque conjugate to $\theta$ is
\begin{equation}
	\tau_\theta=-\frac{\partial \EC}{\partial\theta},
	\label{eq:torque}
\end{equation}
where $\EC$ is the Casimir energy per unit noncompact volume. 

For the massive field, using Eqs.~\eqref{eq:dfdL} and \eqref{eq:torque}, we obtain
\begin{equation}
	\tau_\theta
	=-\mathscr{C}_3 V_p\,m(2m)^2\chi u^2
	\sum_{\ellv\ne0}
	\frac{X_\ell Y_\ell}{L_\ell^{3}}
	K_{3}(mL_\ell).
	\label{eq:torque_massive}
\end{equation}
For the massless field the corresponding expression is
\begin{equation}
	\tau_\theta
	=-\frac{2\Gamma(2)V_p}{\pi^2\sqrt{1-\lambda u^2}}
	\sum_{\ellv\ne0}
	\frac{\chi u^2X_\ell Y_\ell}{L_\ell^{6}}.
	\label{eq:torque_massless}
\end{equation}
The torque vanishes in the Lorentz-invariant limit $\lambda\to0$. For a square compactification its angular dependence is constrained by the discrete fourfold lattice symmetry; for a genuinely rectangular compactification a nonzero orientation-dependent torque is generically allowed. 

\begin{figure}[t]
	\centering
	\includegraphics[width=0.78\linewidth]{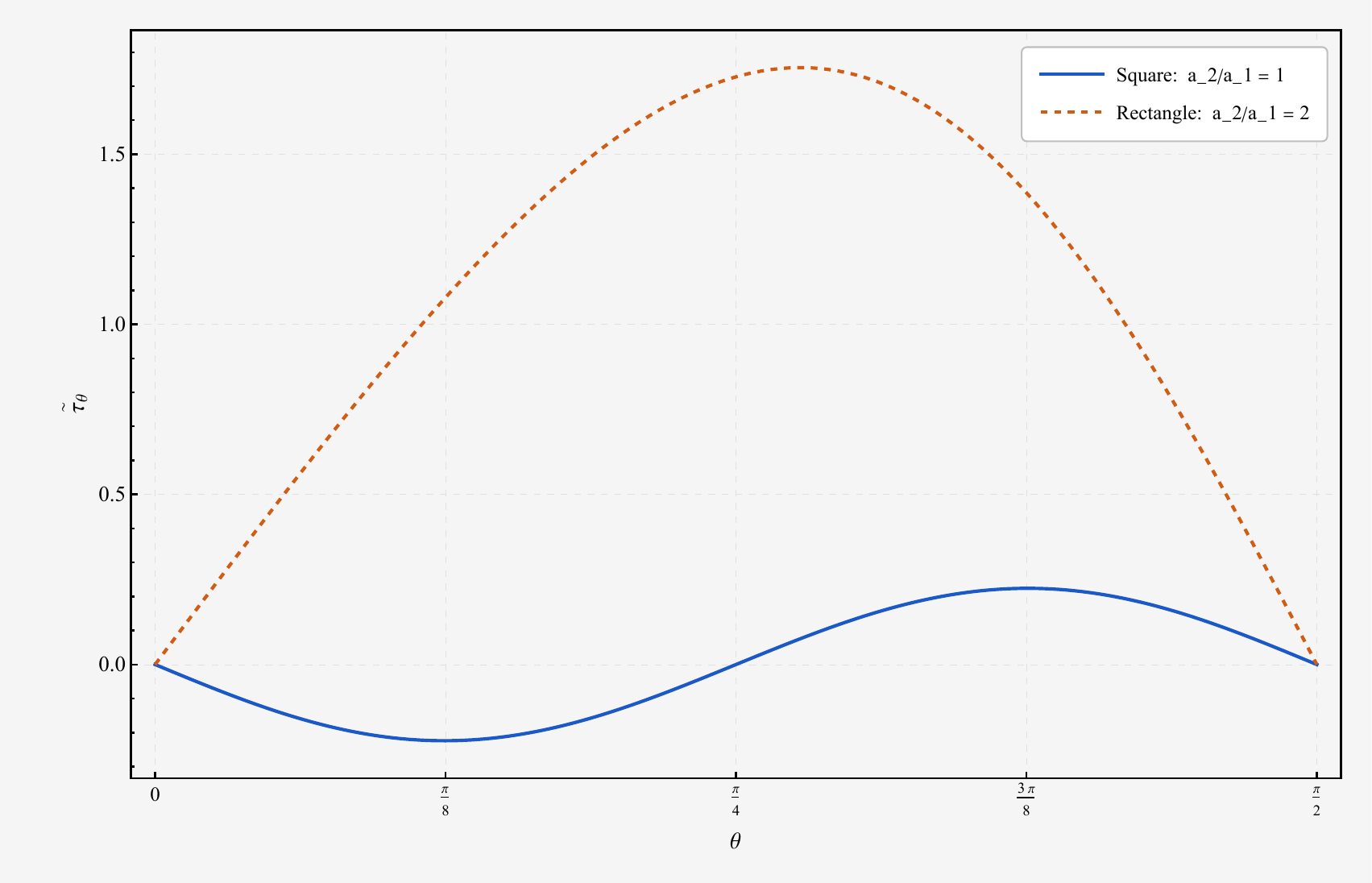}
	\caption{The dimensionless massless Casimir torque $\widetilde{\tau}_\theta=\pi^2 a_1^4\tau_\theta/(V_p\chi u^2)$ for $d=3$, $p=2$, and $\lambda u^2=0.1$, plotted against the orientation angle $\theta$. The curves compare the square lattice ($a_2/a_1=1$) and the rectangular lattice ($a_2/a_1=2$).}
	\label{fig:torque}
\end{figure}

To present the angular dependence in dimensionless form, we define
\begin{equation}
	\widetilde{\tau}_{\theta}
	\equiv
	\frac{\pi^2 a_1^4}{V_p\,\chi u^2}\,
	\tau_{\theta},
	\qquad
	\chi=\frac{\lambda}{1-\lambda u^2}.
	\label{eq:dimensionless_torque}
\end{equation}
This normalization is used for nonzero $\lambda u^2$.
The angular dependence of the normalized massless torque is shown in Fig.~\ref{fig:torque}.

\begin{figure}[t]
	\centering
	\includegraphics[width=0.78\linewidth]{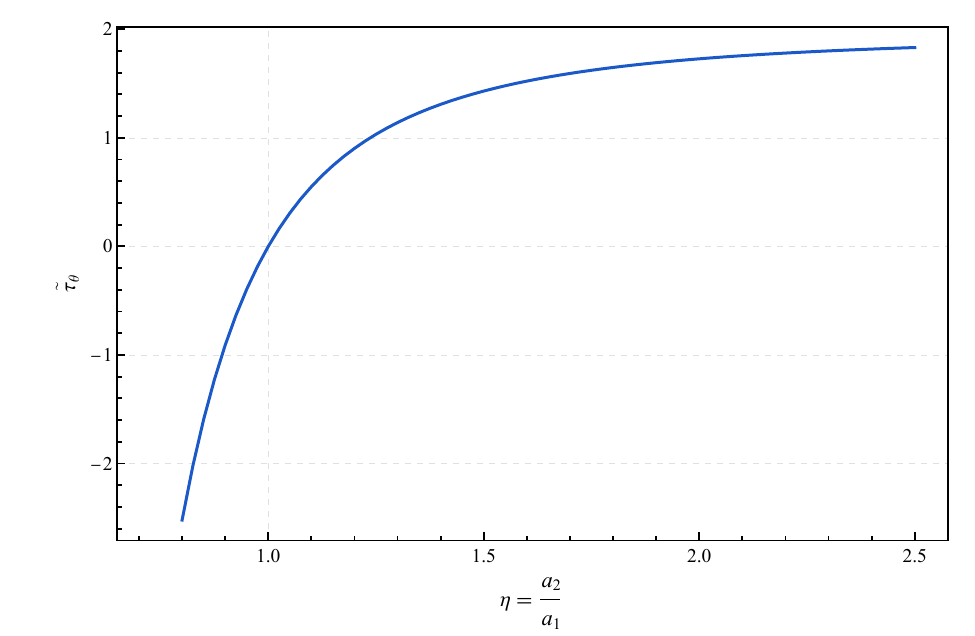}
	\caption{Dimensionless massless Casimir torque $\widetilde{\tau}_\theta=\pi^2 a_1^4\tau_\theta/(V_p\chi u^2)$ at fixed $\theta=\pi/4$, plotted against the aspect ratio $\eta=a_2/a_1$ for $d=3$, $p=2$, and $\lambda u^2=0.1$. The calculation uses $a_1$ as the reference length and a symmetric lattice cutoff $|\ell_i|\le100$. The zero at $\eta=1$ follows from square-lattice symmetry.}
	\label{fig:torque_aspect}
\end{figure}

To isolate the dependence on shape anisotropy, Fig.~\ref{fig:torque_aspect} fixes the relative orientation at $\theta=\pi/4$ and varies the aspect ratio $\eta=a_2/a_1$. The dimensionless torque vanishes at $\eta=1$ by square-lattice symmetry and changes sign as the longer side switches from the $a_1$ to the $a_2$ direction. The lattice sum is truncated symmetrically at $|\ell_i|\le100$; increasing the cutoff from $60$ to $100$ changes the plotted endpoint values by less than $10^{-4}$ for the stated range.

From the perspective of fundamental compactifications, the orientation dependence derived here may be relevant when the relative orientation between a Lorentz-violating background and the compactification geometry is promoted to a dynamical degree of freedom. In such an extension, the Casimir energy would contribute to the effective potential governing that degree of freedom. If $\EC$ denotes the energy per unit volume of the noncompact directions, the quantity $\tau_\theta=-\partial \EC/\partial\theta$ is a generalized Casimir torque density conjugate to the relative orientation. A dynamical stabilization mechanism, however, would require a stationary point of the full effective potential and a separate stability analysis. Flux compactifications provide broader examples in which effective potentials can stabilize compactification moduli \cite{GiddingsKachruPolchinski2002}, but the present calculation does not by itself establish stabilization of the orientation considered here.

\subsection{Consistency check between the massive and massless torques}\label{subsec:torque-consistency}
The massless expression in Eq.~\eqref{eq:torque_massless} can be independently obtained from the massive result in Eq.~\eqref{eq:torque_massive} using
\begin{equation}
    K_3(z)\sim 8z^{-3},\qquad z\to0^+.
\end{equation}
Indeed, $m(2m)^2K_3(mL_\ell)\to 32L_\ell^{-3}$, which yields Eq.~\eqref{eq:torque_massless} with $\mathscr{C}_3=1/(16\pi^2\sqrt{1-\lambda u^2})$. This check fixes both the power of $L_\ell$ and the numerical coefficient in the massless torque.

\section{Massless limit and consistency checks}\label{sec:massless-checks}
The massless limit follows from
\begin{equation}
    K_\nu(z)\sim 2^{\nu-1}\Gamma(\nu)z^{-\nu},
    \qquad z\to0^+.
\end{equation}
Equation~\eqref{eq:master} then gives
\begin{equation}
    {
    \ECsub{m=0}
    =-\frac{\Gamma\!\left(\frac{d+1}{2}\right)}{2\pi^{(d+1)/2}}
    \frac{\V}{\sqrt{1+\lambda u_\mu u^\mu}}
    \sum_{\ellv\in\Z^p\setminus\{0\}}
    \frac{1}{L_\ell^{d+1}}
    .}
    \label{eq:massless}
\end{equation}
This is a generalized Epstein-type lattice sum. Because $p\le d$, the massless sum is convergent.

Several limits provide useful normalization checks. First, for $d=p=1$ and $m=0$,
\begin{equation}
    \ECzero=-\frac{\pi}{6a_1},
    \label{eq:check1d}
\end{equation}
which is the standard periodic massless scalar result in $(1+1)$ dimensions. Second, for $d=3$, $p=1$, and $m=0$,
\begin{equation}
    \ECzero=-\frac{\pi^2}{90a_1^3},
    \label{eq:check3d}
\end{equation}
where $\ECzero$ is already the energy per unit transverse area, consistent with the definition of $\EC$ used throughout the paper. Third, setting $\lambda=0$ in Eq.~\eqref{eq:master} immediately restores the Lorentz-invariant rectangular compactification result.

For a massive field, the asymptotic behavior follows from the large-argument expansion of the modified Bessel function,
\begin{equation}
    K_\nu(z)\sim\sqrt{\frac{\pi}{2z}}e^{-z}
    \left[1+\mathcal O(z^{-1})\right],
    \qquad z\to\infty,
\end{equation}
so the finite-size contribution is exponentially suppressed when $mL_\ell\gg1$. Thus the mass dependence retains the usual decoupling behavior, while LV modifies the geometric lengths entering the exponential factor.

\subsection{Relation to the Dirichlet rectangular-waveguide problem}\label{subsec:dirichlet-comparison}
The periodic compactification studied here is spectrally distinct from a rectangular Dirichlet waveguide. A recent analysis considered the same CPT-even aether-like scalar structure for a rectangular cavity with Dirichlet conditions, focusing on representative coordinate-aligned backgrounds and separable mode spectra \cite{CruzBezerraMartin2026}. Our problem instead uses periodic spatial identification and allows an arbitrary orientation of the background relative to an anisotropic compactification lattice. The two calculations are therefore complementary: the Dirichlet problem probes boundary-induced mode quantization, whereas the present formulation isolates how Lorentz violation deforms the compactification lattice itself. Their final Casimir energies should not be identified term by term.

\section{Discussion}\label{sec:discussion}
The central geometric content of our result is that Lorentz violation deforms the effective metric sampled by the compactification lattice. The spatial kinetic operator is governed by the anisotropic quadratic form $\kk^{\mathsf T}\Mmat\kk$. After the noncompact spatial momenta are integrated out, the compact sector is governed by the Schur complement $\Schur$, which determines the effective squared lengths of the image lattice,
\begin{equation}
	L_{\ell}^{2} = \ellv^{\mathsf T}R\Schur^{-1}R\ellv.
\end{equation}
In the Lorentz-invariant limit, the standard Ambj\o rn--Wolfram image-lattice representation is recovered, whereas for nonzero $\lambda$, the deformed lattice metric generates orientation-dependent spectral lengths and, consequently, an orientation-dependent Casimir energy. The distinction from the effective-metric formulation of Ref.~\cite{Escobar2026} lies in the geometric setting and the role of the reduced quadratic form: here, it determines the image-lattice geometry of a rectangular periodic compactification, whereas Ref.~\cite{Escobar2026} develops the corresponding structure for fluctuation sectors in plate geometries. This geometric viewpoint unifies the axis-aligned limits and the generic-orientation corrections obtained in this work, and provides the basis for the associated generalized Casimir torque.

This perspective clarifies why the familiar special cases look qualitatively different. A purely timelike background leaves the spatial quadratic form unchanged and therefore modifies only the normalization. A spatial background with no compact projection changes only the noncompact integration measure and again yields a multiplicative factor. A background with a compact projection changes the effective compact lengths, and a generic orientation generates a non-diagonal quadratic form in the rectangular lattice basis.

The square-to-rectangle comparison in Fig.~\ref{fig:angular}, together with the fixed-angle aspect-ratio scan in Fig.~\ref{fig:torque_aspect}, gives a direct physical manifestation of this distinction. For $a_1=a_2$ the square-lattice symmetry suppresses the leading angular modulation, whereas for $a_1\ne a_2$ the vacuum energy responds to the orientation of the preferred direction and the conjugate torque is generically nonzero. Hence the effect of Lorentz violation is not characterized by a single universal multiplicative coefficient: the geometry of the finite-size sector itself can carry information about the background orientation.

An additional consistency point concerns earlier calculations in which the background was chosen along one noncompact coordinate. Because only the momentum component parallel to that background is rescaled, the corresponding Jacobian from the continuous momentum integral is proportional to a single square-root factor, independently of the total number of noncompact directions. The present block-matrix formulation makes this immediately manifest. This observation is useful for separating a genuine dimensional dependence of the Casimir kernel from a Jacobian produced by a one-component change of variables.

Although the numerical illustrations in this work employ a relatively large deformation parameter, such as $\lambda |\mathbf u_s|^2=0.1$ in purely spacelike configurations, these values are intended to make the geometric modifications clearly visible and should be regarded as illustrative rather than phenomenologically motivated. Experimental constraints on Lorentz violation are stringent, but their interpretation depends on the operator, field sector, and combinations of coefficients under consideration. In particular, bounds on coefficients in the Standard-Model Extension (SME) cannot be translated directly into a model-independent bound on $\lambda u^\mu u^\nu$ without specifying the normalization of the background vector and establishing the appropriate matching to the relevant SME coefficients \cite{KosteleckyRussell2011}.

For small Lorentz-violating coefficients, the leading correction to the Casimir energy contains an orientation-independent contribution from the kinetic normalization and, whenever the background has a compact projection, a geometry-dependent contribution from the deformation of the image-lattice distances. The corresponding generalized Casimir torque is likewise controlled by the orientation-dependent part of the energy. A quantitative assessment of experimental detectability would require evaluating the relevant force or torque observable in a physically realizable configuration and comparing the predicted signal with the uncertainties of the corresponding measurement. The present calculation alone does not establish a direct experimental sensitivity bound, since it concerns a free scalar field on a periodic compactification rather than the electromagnetic field between material boundaries in conventional laboratory Casimir experiments.

Condensed-matter analogs and structured photonic systems may provide settings in which some aspects of the orientation-dependent dispersion studied here can be explored. Tilted Weyl cones in type-II Weyl semimetals and engineered anisotropic or hyperbolic photonic dispersions offer examples of effective spectra with preferred directions \cite{Soluyanov2015,Hu2020}. These analogies are primarily kinematic: the cited systems do not, by themselves, realize the scalar-field Lagrangian or the rectangular periodic compactification considered in this work. Establishing a direct experimental realization would require a platform-specific effective Hamiltonian, an appropriate implementation of the boundary conditions, and a calculation of the relevant vacuum or quasiparticle contribution to the mechanical response. The generalized torque derived here therefore provides a theoretical motivation for such investigations rather than a direct prediction of a measurable torque in an existing material platform.

The present model remains intentionally restricted. We consider a free massive scalar, a constant background vector, flat spacetime, zero temperature, and periodic spatial compactification. We do not include dynamical aether fields, higher-derivative Lorentz-violating operators, interactions, curved backgrounds, finite-temperature effects, or generic Dirichlet boundaries with arbitrary background orientation. These extensions would require additional spectral or renormalization input and should not be inferred from Eq.~\eqref{eq:master}. Analog condensed-matter systems with engineered effective dispersions may provide a setting for exploring related effects, but quantitative predictions would require a consistent mapping between the effective material parameters and the field-theoretic coefficients used here. Cosmological or early-universe applications would likewise require a separate physical realization of the compactification and Lorentz-violating background.

\section{Conclusion}\label{sec:conclusion}
We have derived a general all-dimensional expression for the Casimir energy of a free massive scalar field on the rectangular compactification $\R^{1,r}\times T^p$ in the presence of a constant CPT-even aether-like Lorentz-violating background. The calculation reorganizes the problem in terms of an effective spatial kinetic matrix and its Schur complement with respect to the noncompact directions.

The renormalized energy is given by the rapidly convergent Bessel lattice sum in Eq.~\eqref{eq:master}. The result is exact in the LV parameter within the stability domain of the quadratic theory. Its structure separates the LV correction into an invariant normalization factor, $[1+\lambda u_\mu u^\mu]^{-1/2}$, and an orientation-dependent deformation of the Poisson-resummed image lattice through $L_\ell$.

The timelike, compact-axis, and noncompact-axis configurations are recovered as immediate limits. The generic-orientation formula additionally predicts angular dependence of the vacuum energy whenever the background has a compact projection and the compactification lattice is geometrically anisotropic. The corresponding compact-direction pressures are given explicitly in Eq.~\eqref{eq:explicitpressure}, and the two-dimensional angular example leads to the Casimir torque in Eq.~\eqref{eq:torque}.

The primary contribution of this work is the exact, all-dimensional spectral analysis of a rectangular periodic compactification in the presence of a Lorentz-violating background with an arbitrary spatial orientation. By systematically identifying how the anisotropic background deforms the effective image-lattice metric, this formulation isolates the purely geometric response of the quantum vacuum from boundary-induced finite-size effects. Overall, these results demonstrate that quantum zero-point fluctuations in compactified topologies encode both the magnitude and the directional orientation of Lorentz symmetry violation through explicit geometric forces and generalized Casimir torques.

\section*{Statements and Declarations}
\paragraph{Competing interests.}
The authors declare no competing interests directly related to this work.

\paragraph{Data availability.}
No datasets were generated or analyzed during the current study. The numerical illustrations in Figs.~\ref{fig:masslambda}--\ref{fig:torque_aspect} are reproducible directly from the analytical expressions given in Eqs.~\eqref{eq:master}, \eqref{eq:massless}, \eqref{eq:torque_massless}, and~\eqref{eq:dimensionless_torque}.

\appendix
\section{Technical derivation of the Gaussian and Poisson factors}\label{app:technical-derivation}
For completeness, we record the main algebraic steps leading from Eq.~\eqref{eq:formalE} to Eq.~\eqref{eq:master}. With the block decomposition in Eq.~\eqref{eq:blockM}, completing the square gives Eq.~\eqref{eq:schur}. The change of noncompact variables
\begin{equation}
    \qq_n=\kk_n+C^{-1}B^{\mathsf T}\kk_c
\end{equation}
leaves the measure invariant and yields
\begin{equation}
    \int\frac{\dd^r q_n}{(2\pi)^r}
    e^{-t\qq_n^{\mathsf T}C\qq_n}
    =\frac{1}{(4\pi t)^{r/2}\sqrt{\det C}}.
    \label{eq:gaussian}
\end{equation}
The compact exponential is
\begin{equation}
    \exp\!\left[-4\pi^2t\,\nn^{\mathsf T}R^{-1}\Schur R^{-1}\nn\right],
\end{equation}
and its Poisson transform is Eq.~\eqref{eq:poisson}. The determinant identity
\begin{equation}
    \det\Mmat=\det C\,\det\Schur
\end{equation}
then combines the noncompact Gaussian determinant and the compact Poisson determinant into the single factor $1/\sqrt{\det\Mmat}$. Finally, the integral identity Eq.~\eqref{eq:besselidentity} produces the modified Bessel function in Eq.~\eqref{eq:master}.

\end{document}